\documentclass[10pt,journal,compsoc]{IEEEtran}
\usepackage[hidelinks]{hyperref}
\usepackage{cite}

\usepackage{comment}

\ifCLASSINFOpdf
\else
\fi
\usepackage[normalem]{ulem}
\usepackage{amsmath,amssymb,amsfonts}
\usepackage{algorithmic}
\usepackage{graphicx}
\usepackage{textcomp}
\usepackage{xcolor}
\usepackage{listings}
\usepackage{fancyvrb}
\usepackage{framed}
\usepackage{multirow}
\usepackage{multicol}
\usepackage[linesnumbered, ruled]{algorithm2e}
\usepackage{stmaryrd}
 \usepackage{ulem} 

\definecolor{hookersgreen}{rgb}{0.0, 0.44, 0.0}
\definecolor{amethyst}{rgb}{0.6, 0.4, 0.8}
\definecolor{antiquefuchsia}{rgb}{0.57, 0.36, 0.51}
\definecolor{azure}{rgb}{0.0, 0.5, 1.0}
\definecolor{blue-violet}{rgb}{0.54, 0.17, 0.89}

\SetCommentSty{mycommfont}

\makeatletter
\newcommand{\multiline}[1]{%
  \begin{tabularx}{\dimexpr\linewidth-\ALG@thistlm}[t]{@{}X@{}}
    #1
  \end{tabularx}
}
\makeatother

\newcommand{\vic}[1]{\textcolor{black}{#1}}
\newcommand{\ivan}[1]{\textcolor{black}{#1}}
\newcommand{\die}[1]{\textcolor{black}{#1}}
\newcommand{\tse}[1]{\textcolor{black}{#1}}

\begin{document}

%
\title{Dynamic Slicing by On-demand Re-execution}
%
%
%
\author{Ivan~Postolski, Victor~Braberman, Diego~Garbervetsky and Sebastian~Uchitel
\IEEEcompsocitemizethanks{\IEEEcompsocthanksitem Ivan~Postolski, Victor~Braberman, Diego~Garbervetsky and Sebastian~Uchitel are with UBA, ICC, CONICET. Argentina. Email: \{ipostolski, vbraber, diegog, suchitel\}@dc.uba.ar
\IEEEcompsocthanksitem Sebastian~Uchitel is with Imperial College London. UK. Email: s.uchitel@imperial.ac.uk
\IEEEcompsocthanksitem Corresponding Author: Víctor~Braberman.
}
}

\IEEEtitleabstractindextext{
\begin{abstract}
In this paper, we propose a novel approach that aims to offer an alternative to the prevalent paradigm to dynamic slicing construction. Dynamic slicing requires dynamic data and control dependencies that arise in an execution. During a single execution, memory reference information is recorded and then traversed to extract dependencies. Execute-once approaches and tools are challenged even by executions of moderate size of simple and short programs. We propose to shift practical time complexity from execution size to slice size. In particular, our approach executes the program multiple times while tracking targeted information at each execution.
We present a concrete algorithm that follows an on-demand re-execution paradigm that uses a novel concept of frontier dependency to incrementally build a dynamic slice. To focus dependency tracking, the algorithm relies on static analysis. We show results of an evaluation on the SV-COMP benchmark and Antrl4 unit tests that provide evidence that on-demand re-execution can provide performance gains particularly when slice size is small and execution size is large.
\end{abstract}

}

\maketitle

\IEEEdisplaynontitleabstractindextext

%
\IEEEpeerreviewmaketitle

\section{Introduction}
The concept of dynamic program slicing was proposed in the early 90s by Agrawal et.al. \cite{agrawal1990dynamic} and Korel et.al. \cite{korel1988dynamic} as an alternative to static slicing. Static slicing suffers from imprecision (too many statements) resulting from considering too many execution paths. However, in some programming tasks such as debugging or testing interest is only in one specific execution path that, for instance, triggers a system failure. Korel and Agrawal \cite{korel1988dynamic, agrawal1990dynamic} then proposed to solve a different problem: given an execution path (e.g., a test), find all the statements that affect program values at a point of interest (e.g.,  a test assertion); namely a dynamic slice.
\tse{Since then, dynamic slicing has been proposed as a backend technique for debugging~\cite{li2020more}, fault localization~\cite{soremekun2021locating}, and automated program repair~\cite{guo2018empirical}, amongst others.}

\tse{State-of-the-art dynamic slicing techniques struggle in terms of efficiency and scale. Algorithms \textit{register}
memory reference information \cite{zhang2003precise,wang2008dynamic}
from a single program execution (i.e., \textit{execute-once}), and afterwards
\textit{traverse} this information 
to extract dependencies that can be used to produce a 
slice. Scalability of these techniques is known to suffer when execution size, the number of executed instructions, grows \cite{zhang2005cost,zhang2006dynamic,wang2014drdebug}.}


Given that dynamic slices are known to be small in size~\cite{binkley2004survey}, our hypothesis is that dynamic slicing computation time
can be improved by detecting relevant dependencies at execution time rather than registering and traversing memory references. This may lead to a slicing process that is more dependent on slice size and less on execution size.

We propose an approach that incrementally builds a slice by acquiring targeted information about dependencies through repeated execution of the same program but varying instrumentation based on the statements already identified as part of the slice. \tse{We refer to this approach as dynamic slicing by \textit{on-demand re-execution}.}

Concretely, we calculate the slice set $S$ by starting with a set containing the target statement (commonly called the slicing criteria). In each iteration we obtain
--using a static dependency analysis--
the potential \ivan{direct} data dependencies from statements in $S$ to statements not in $S$, we call these dependencies \textit{frontiers}. Of course, many of such frontiers may not be executed (i.e., the def-use pair is not covered~\cite{su2017survey,santelices2007efficiently}) due to the approximate nature of the static analysis. Thus, we instrument and execute the program to check which of those frontiers are actually executed. If a frontier execution is corroborated --i.e., we detect that a node not included in $S$ writes a value that is read by a node in $S$--
we add the observed writing node to the slice and re-iterate. Naturally, the computation stops when no new frontier executions are corroborated.
The algorithm includes careful and efficient handling of calling contexts to support detection of frontier dependency executions incrementally in an inter-procedural setting.

\tse{Our evaluation looks at comparing \textit{on-demand re-execution} and \textit{execute-once} paradigms in terms of overall performance and how gains and losses are affected by execution and slice size.} Specifically we compare against Javaslicer~\cite{javaslicer} and Slicer4J~\cite{ahmed2021slicer4j}.

We developed an implementation that targets the Java programming language implementing a static dependency analysis using  CodeQL~\cite{codeql,de2007ql}, and a dynamic instrumentation for frontier tracking, avoiding a re-compilation task per re-execution.
We split our evaluation between a controlled experimentation in an environment where subjects are classic algorithms implemented in Java (SV-COMP benchmark~\cite{SV-COMP}), and an industrial feasibility study in a open source, widely adopted subject, Antlr4\cite{antlr4}.

For the SV-COMP benchmark the main results on the construction of more than 1700 slices --unique combinations of inputs and target statements-- include:
$i)$ an average of 124x gain in time when compared with a state-of-the-art tool~\cite{javaslicer} that implements a \tse{execute-once} strategy, observing that in general gains are typically almost linear to the execution size;
 and $ii)$ a statistically significant negative Spearman correlation coefficient between slice size and gains with $p<0.001$ for groups of slices built with the same input.

On the other hand, in the Antlr4 case study, we report results for the construction of 50 slices from real unit tests assertions where we obtain: $i)$ an average 8x gain in overall slice time. Total time by the state-of-the-art tool was 70min compared to a little more than 8min of our approach, $ii)$ a statistically significant negative Spearman correlation coefficient $p<0.001$ between the produced slice size and gains of our approach versus available \tse{execute-once} tooling.

Finally, for the Antlr4 case study, considering its only available performance test, we found that state-of-the-art tooling ran out of memory in 15min while processing~3GB traces while our approach was able to slice them in 5min.

Results also provide evidence that our on-demand incremental dependency corroboration plays a key role in our approach performance and its shift in practical complexity to slice size rather than execution size.

To summarize, the contributions of this paper are $i)$ a novel idea, on-demand re-execution, that challenges the execute-once strategy that is predominant in dynamic slicing, $ii)$ an algorithm for on-demand re-execution dynamic slicing based on static frontier dependency analysis, and $iii)$ an evaluation that provides evidence that on-demand re-execution can provide performance gains for dynamic slice computation particularly when the slice size is small and the execution size large.

\section{Background}

Korel-Laski's seminal work defines dynamic slicing by instantiating a program's structure with a \die{control} flow graph $G=(\mathcal{N},\mathcal{A},s,e)$ where 1) $\mathcal{N}$ is a set of nodes, 2) $A$ is a set of arcs (a binary relation on $\mathcal{N}$), and 3) $s$, and $e$, are respectively unique entry and exit nodes. A node corresponds to a statement in the program, and an arc $(n,m)$ is in $\mathcal{A}$ when a potential transfer of control exists between $n$ and $m$.


A path $p$ from the entry node $s$ to some node $n$ in $\mathcal{N}$, is a sequence $\langle s,..., n \rangle$ of nodes, of length $k$, such that $(p[i], p[i+1])$ in $\mathcal{A}$, for all $1 \leq i < k$. A path that has been executed for some input $x$ will be referred to as an execution trace $T_x$. Since a node can be executed multiple times in a trace, nodes are distinguished in an execution trace by their occurrence number. More formally, given a trace $T_x$ a node occurrence $n_w^q$ represents the $q$-th appearance of a node $n_w$ in $T_x$. Also, given a trace $T_x$, $(n_w^p, n_r^t)$ is in $DU_{T_x}$ whenever an \die{occurrence of a node} $n_w^p$ defines a memory address that is later used by \die{an occurrence} $n_r^t$ without being redefined by other node in $T_x$ between $n_w^p$ and $n_r^t$. For simplicity we note $n_w^q < n_r^t$ whenever a node $n_w^q$ occurs in the trace $T_x$ before $n_r^t$.

Given a trace $T_x$ and a node occurrence $I^q$ present in $T_x$, Korel-Laski's slicing \cite{korel1988dynamic} is a fixed-point algorithm that builds the smallest set of nodes such that ($a$) it contains node occurrence $I^q$, and ($b$) it is closed under: data dependencies for all variables, control dependencies, and the identity relation. In a few words, when a node occurrence $n_r^t$ happens to be in the computed set of relevant occurrences, then any $n_w^p$ such that $(n_w^p, n_r^t)$ is in $DU_{T_x}$
is also part of the set of nodes. As well as any $n_c^p$ nodes that control the execution of $n_r^t$ - as defined by Ferrante et al.~\cite{ferrante1987program}- where $p$ is the last occurrence of $n_c$ before $n_r^t$ , and -by the identity relation- any occurrence $n_r^p$ where $p < t$.

\section{On-demand re-execution slicing}

Our approach, which aims at corroborating relevant dependencies, requires some information that can be statically obtained plus some standard run-time capabilities for instrumentation and detection.



We assume that given a node $n_r$ it is possible to statically compute the set of nodes tuples $\{(n_{w0}, n_r),...,(n_{wk}, n_r)\}$, that we refer to as $SD_{data}(n_r)$ or just static \die{data} dependencies, where $n_{wi}$ is a node that may write an address in memory later used (read) by $n_r$. We assume the set of nodes $\{n_c \, | \, n_c \text{ controls } n_r\}$, namely $SD_{control}(n_r)$, where controls is the classic control-dependency relation defined by Ferrante et al.~\cite{ferrante1987program}. As many static analyses can be adapted to output such both sets of dependencies, we generically refer to the analysis that computes them as $SDA$.
In this paper, we use static taint analysis for such purpose~\cite{codeql,de2007ql}. On the other hand, we assume runtime support is able to pinpoint memory addresses used or defined by a program node execution. Like in def-use coverage analysis \cite{su2017survey} and dynamic tainting \cite{kang2011dta}, that support is instrumental for our targeted dependency-corroboration analysis we introduce in the next sections.




\subsection{Intra-procedural Algorithm}

To introduce  the approach in its simplest form, we present an intraprocedural version of the on-demand re-execution slicing pseudo-code (Algorithm \ref{intradfrontierslice}). Next section will show the full version that also handles procedure/method calls.

Our slice formulation takes a node occurrence, $I^q$, as the criterion and targets all the variables used by it, the input of the program to be sliced, $x$, and the $\mathit{SDA}$ as parameters.


\begin{algorithm}
   \algsetup{linenosize=\tiny}
  \footnotesize

Procedure \textbf{OnDemandIntraProceduralSlice} ($x,I^q,SDA$):
   \caption{}
    \label{intradfrontierslice}
    \tcp{Variable declarations:}
    Let $S$ be a set of program nodes;\\
    Let $\mathit{obs\_defs}$ and $\Delta{\mathit{control}}$ be a set of  nodes;\\ 
    Let $\mathit{checked\_frontiers}$ be a set of tuples of nodes;\\
    Let $\mathit{frontiers2check}$ be a set of tuples of nodes;\\

    \tcp{Sub-routines definitions:}
    Let $\mathit{frontiers (S)}$ $=_{def}$
    $\{(n_w,n_r) \: | \: n_w \in  SD_{data}(n_r) \land n_r \in S \land n_w \not\in S\}$\\

    Let $ \mathit{control}(n) =_{def} \{ n' \: | \text{ where } n' \in SD_{control}(n)\}$ \\


    Let $\mathit{missing\_control}(S)=_{def}$
    $\{n' \: | \: \exists n, n'.  n \in S \land n' \not\in S  \land n' \in \mathit{control}(n) \}$


        Let $\mathit{exercised\_frontiers}(f2c,x,I^q):=_{def}$
    $\{(n_w,n_r)|\: \exists n_w,n_r,k, p. \: (n_w,n_r) \in f2c \: \land  (n_w^k,n_r^p) \in DU_{T_x} \land n_r^p < I^q\}$




    \tcp{Main algorithm:}
    $S$ = $\{I\} \cup \mathit{control}(I)$;

    $\mathit{checked\_frontiers}$ = $\varnothing$;

    $\Delta{\mathit{control}} = \varnothing$;

\Repeat{$\mathit{obs\_defs} = \varnothing \land \Delta{\mathit{control}} = \varnothing $ }{

      $\mathit{frontiers2check}=\mathit{frontiers}(S) - \mathit{checked\_frontiers}$;

      $\mathit{obs\_defs}$ = $\{n_w \:|\: \exists n_w,n_r. \: (n_w,n_r) \in exercised\_frontiers(\mathit{frontiers2check},x,I^q)\}$


     $S$ = $S \cup \mathit{obs\_defs}$;

     $\Delta{\mathit{control}} = \mathit{missing\_control}(S)$;

     $S = S \cup \Delta{\mathit{control}}$;

     $\mathit{checked\_frontiers}$ = $\mathit{checked\_frontiers} \cup \mathit{frontiers}(S)$;

}

\KwRet{$S$}\;

\end{algorithm}
We begin our slicing algorithm by adding to $S$ the criteria node $I$ --line 10-- and $I$'s control dependencies; and initiating $\mathit{checked\_frontiers}$ and $\Delta{\mathit{control}}$ as empty sets --line 11,12--.

Then we perform an on-demand re-execution cycle --line 14 to 19--, this fixed-point loop performs five operations: it calculates the frontiers to check of $S$ --line 14-- it executes an instrumented version of the program to corroborate which of those frontiers are actually exercised and filters its detected definition nodes --line 15--, then it adds these observed nodes to the current slice $S$ --line 16--, adds the missing control dependencies of nodes currently in $S$ --lines 17,18--, and updates the already checked frontiers --line 19--. Finally, the routine verifies whether the observed definitions are empty (i.e., none of the frontiers were exercised) and also whether there are no more missing control dependencies. If this is the case then the algorithm terminates returning the $S$ set.

The algorithm always terminates, as every iteration of the loop adds elements to $S$, which is bounded by the size of the program being sliced. The post condition is that for all nodes in $S$ there are no dependencies to other nodes outside $S$ when executing $x$.

Note that, the frontier corroboration --line 15-- is done (on-demand) for frontiers of $S$ --that vary as $S$ is extended-- without the frontiers checked in previous iterations. This avoids unnecessary instrumentation, as once a frontier is analyzed in a re-execution step whether it is exercised or not will not change in the next iterations.

In a few words, the algorithm can be also regarded as a fixed-point iteration that computes the closure of (static) dependencies with a key dynamic analysis step that filters out those static dependencies that are spurious because they are not exercised by the execution flow or because they are actually spurious dependencies due to \die{static analysis approximations (e.g., alias/points-to analysis, control flow, etc).}

\vic{The number of frontiers to check at each iteration depends on the data dependency structure of the program and the precision of the underlying static analysis. In any case, and considering all iterations, the number of edges of the static slice constitutes an upper bound for the number of frontiers that are checked.}

\subsubsection{Detecting exercised frontiers}

Detection can be done by executing (or replaying) an instrumented version of the program. Our instrumentation consists in tracking into a shadow memory $M'$ the definitions performed by nodes that the static analysis detects as potential writers of variables read by readers of the frontiers to be checked.  That is, each time one of those potential writers $w$ defines a memory address $m$, $M'$ at $m$ is annotated with $w$. When a frontier reader node is executed, the instrumentation logic checks $M'$ at the used addresses and it adds those last writers that happen to be frontier writers (that is, there is evidence of frontier being exercised). Note that only current frontier readers and the potential writers need to be treated by the instrumentation. \vic{Moreover, the memory consumed is in direct relationship with the memory locations actually defined by writer nodes (which in turn, assuming sound static dependency analysis, is a superset of memory locations used by readers). The memory required could be in practice, substantially less than the total memory space defined by the program. }

\subsection{Inter-procedural algorithm}
\label{inter-section}


\begin{figure}[h]
    \centering
    \includegraphics[scale=1.5]{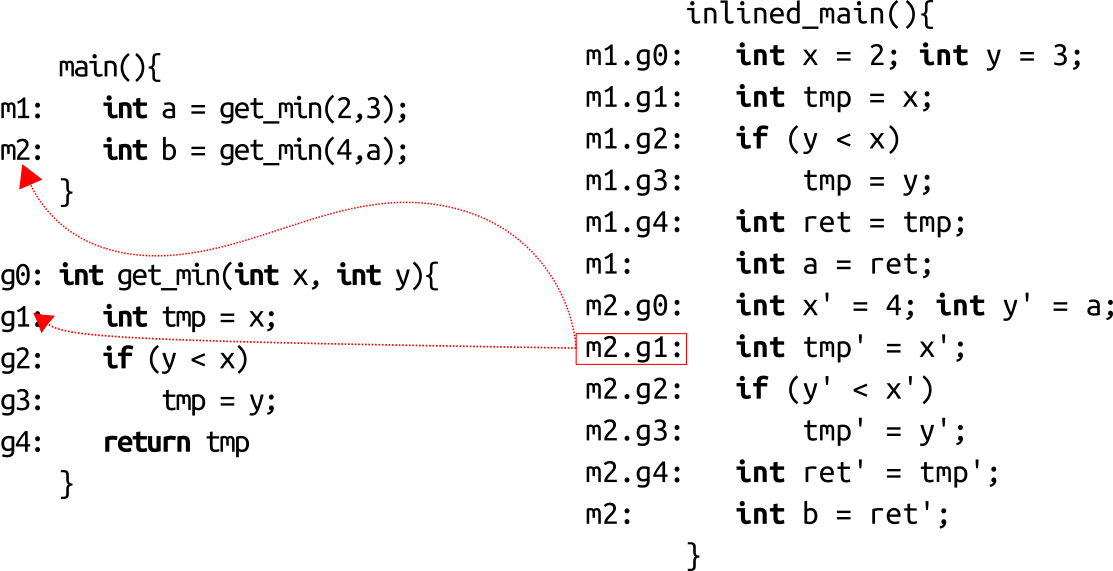}
    \caption{Inter-procedural code example with the inlined program and node mapping}
    \label{inter-example}
\end{figure}

\begin{table}[]
\centering
\caption{Intra-procedural algorithm run for the $\mathtt{inlined\_main()}$ program (Figure~\ref{inter-example}) with node $\mathtt{m2}$ as criterion.}
\label{intra-inline}
\begin{tabular}{|c|l|l|}
\hline
\multicolumn{1}{|l|}{Iter.} & \multicolumn{1}{c|}{$\Delta$$S$} & \multicolumn{1}{c|}{$\mathit{frontiers2check}$} \\ \hline
1 & \{$\mathtt{m2}$\} & \{($\mathtt{m2.g4}$,$\mathtt{m2}$)\} \\ \hline
2 & \{$\mathtt{m2.g4}$\} & \{($\mathtt{m2.g3}$,$\mathtt{m2.g4}$),($\mathtt{m2.g1}$,$\mathtt{m2.g4}$)\} \\ \hline
3 & \{$\mathtt{m2.g3}$\} & \{($\mathtt{m2.g0}$,$\mathtt{m2.g3}$)\} \\ \hline
4 & \{$\mathtt{m2.g0}$\} & \{($\mathtt{m1}$,$\mathtt{m2.g0}$)\} \\ \hline
5 & \{$\mathtt{m1}$\} & \{($\mathtt{m1.g4}$,$\mathtt{m1}$)\} \\ \hline
6 & \{$\mathtt{m1.g4}$\} & \{($\mathtt{m1.g3}$,$\mathtt{m1.g4}$),($\mathtt{m1.g1}$,$\mathtt{m1.g4}$)\} \\ \hline
7 & \{$\mathtt{m1.g1}$\} & \{($\mathtt{m1.g0}$,$\mathtt{m1.g1}$)\} \\ \hline
8 & \{$\mathtt{m1.g0}$\} & $\varnothing$ \\ \hline
\end{tabular}
\end{table}

\begin{table}[]
\centering
\caption{Inter-procedural algorithm run for the $\mathtt{main()}$ program (Figure~\ref{inter-example}) with node $\mathtt{m2}$ as criterion.}
\label{inter-inline}
\begin{tabular}{|c|l|l|l|}
\hline
\multicolumn{1}{|l|}{Iter.} & \multicolumn{1}{c|}{$\Delta$$S$} & \multicolumn{1}{c|}{$\mathit{frontiers2check}$} & \multicolumn{1}{c|}{$\mathit{obs\_fr}$} \\ \hline
1 & \{$\mathtt{m2}$\} & \{($\mathtt{m1}$,$\mathtt{m2}$),($\mathtt{g4}$,$\mathtt{m2}$)\} & \begin{tabular}[c]{@{}l@{}}\{($\mathtt{m1}$,$\varnothing$,$\mathtt{m2},\varnothing$),\\ ($\mathtt{g4}$,\{$\mathtt{m2,g0}$\},$\mathtt{m2,\varnothing}$)\}\end{tabular} \\ \hline
2 & \{$\mathtt{m1,g4,g0}$\} & \{($\mathtt{g3}$,$\mathtt{g4}$),($\mathtt{g1}$,$\mathtt{g4}$)\} & \begin{tabular}[c]{@{}l@{}}\{($\mathtt{g3}$,\{$\mathtt{m2,g0}$\},$\mathtt{g4}$,\{$\mathtt{m2,g0}$\}),\\ ($\mathtt{g1}$,\{$\mathtt{m1,g0}$\},$\mathtt{g4}$,\{$\mathtt{m1,g0}$\})\}\end{tabular} \\ \hline
3 & \{$\mathtt{g3,g2,g1}$\} & $\varnothing$ & $\varnothing$ \\ \hline
\end{tabular}
\end{table}

A naif adaptation of our intra-procedural algorithm to handle inter-procedural programs would begin by removing method calls with succession of inline expansions. Inlining places fresh copies of the nodes of a method body at each call-site location producing a new (inlined) program\footnote{Note that, in theory, an inlining approach would always be possible as we are dealing with finite executions (until the criteria is executed), and a bounded number of inline expansions would suffice to generate an equivalent program.} that has zero method calls, thus our intra-procedural algorithm can be executed for such program. Afterwards, the resulting slice of the inlined program would be easily translated to a slice of the original program by a reverse mapping of the nodes in the inlined program to the corresponding nodes, and call-site nodes, of the original program. We illustrate this in the example code (Figure~\ref{inter-example}).

This naif inlining approach, although sound and precise, unfortunately, would  lead to a growth of the main complexity parameter of our approach: the inlined program (and its slice) would typically grow considerably in size. This would constitute a problem to our approach as more re-executions would be required to slice the inlined program. In the example, the inlined program requires 7 re-executions (Table~\ref{intra-inline}) for a fairly small program.  Moreover, in the presence of recursion, size growth of the original program might closely follow execution size, a complexity parameter we would like to detach from, at least, in practical terms.

Inlining solves the problem of identifying a node calling context by cloning them syntactically per call-site. Instead, our
approach (Algorithm~\ref{interprocedural}) works with the original program but augments the exercised frontiers with the calling contexts at the moment of definition and use of the frontier nodes (\textit{contextualized} frontiers).
Given a $\mathit{frontier2check}$ tuple $(w,r)$, our inter-procedural algorithm detects the contextualized frontiers  $(w,w_{ctx},r,r_{ctx})$ that happen in runtime --line 11--. These are the exercised frontiers along with the \vic{set of their associated} call stack elements (call-site nodes) at the moment of the definition and use respectively.

Then, a contextualized frontier $w$ node is added to the slice only when the reader node $r$ happens in a context $r_{ctx}$ whose call-sites are all included in the current slice $S$ --line 21--. If that is not the case, it is safe to consider that $r$ will not be executed, within the context $r_{ctx}$, by the current slice.

Note that, the algorithm initializes $S$ with the criteria node $I$ --line 12-- and $I$'s control dependencies --line 13--, however as our criteria $I^q$ is indeed an occurrence in a given context, the elements of its call-stack (and their control dependencies) are also added into the initial slice.

On the other hand, call-site nodes in $w_{ctx}$ of a relevant occurrence a frontier's writer node $w$
are then also incorporated into the computed slice --line 22--. This has the effect of the reverse mapping of the inlining approach. However, in comparative terms, it means that many frontiers of the inlining approach could be confirmed in a singled (re-)execution. For instance, in the example our inter-procedural algorithm performs only 2 executions (Table~\ref{inter-inline}) instead of 7.


Finally, as the slice is incrementally built by adding nodes in each iteration --lines 24,25,26--, some of the exercised frontiers that we safely disregarded because $r\_{ctx}$ is not all included in $S$ might actually become relevant in further iterations, thus it would be incorrect to rule these frontiers as checked. Instead we must \textit{keep} them --line 27-- until they \vic{either} become relevant or our algorithm terminates.

Thus, in a few words, the contextualized-frontiers approach identifies potential frontiers using the dependencies detected by the static analysis over the original program, it does not increase the number of lines and frontiers to check like the inlining approach, and leaves the task of identifying relevant occurrences of read and write nodes to the dynamic phase. Yet, the burden introduced to work with contextualized frontiers does not add any factor that might couple the use of time and memory resources to execution size, we expand on this in the following sub-section.

\begin{algorithm}
   \algsetup{linenosize=\tiny}
  \footnotesize
Procedure \textbf{OnDemandInterSlice} ($x,I^{q},SDA$):
    \caption{}
    \label{interprocedural}
    \tcp{Variable declarations:}
    Let $S$ be a set of program nodes;\\
    Let $obs\_defs$ and $obs\_ctxs$ be a set of nodes;\\
    Let $checked\_frontiers$ be a set of tuples of nodes;\\
    Let $kept\_frontiers$ be a set of tuples of nodes;\\
    Let $frontiers2check$ be a set of tuples of nodes;\\
    Let $obs\_fr$ be a set of contextualized tuples of nodes;

    \tcp{Sub-routines definitions:}
    Let $frontiers(S):=_{\mathit{def}}$
    $\{(n_w,n_r)\: | \: \exists n_w,n_r. n_r \in S  \land n_w \in SD_{data}(n_r) \}$

    Let $control(n):=_{\mathit{def}} \{ n' \: | \: \exists n'. n' \in SD_{control}(n)\}$

    Let $missing\_control(S):=_{\mathit{def}}$
 $\{n' \: | \: \exists n,n'.  n \in S \land n' \not\in S \land n' \in control(n) \}$

 Let $\mathit{exercised\_fr\_with\_ctxs}(f2c,x,I^q):=_{def}$
    $\{(w,w_{ctx},r,r_{ctx}) \:|\: \exists w,r,k,p. \: (w,r) \in f2c \: \land  (w^k,r^p) \in DU_{T_x} \land w_{ctx} = ctx(w^k) \land r_{ctx} = ctx(r^p) \land r^p < I^q\}$

    \tcp{Main algorithm:}
    $S$ = $\{I\} \cup ctx(I^{q})$;

    $\Delta{control} = missing\_control(S)$;

    $S = S \cup \Delta{control}$;

    $checked\_frontiers$ = $\varnothing$;

    $kept\_frontiers = \varnothing$;

    $\Delta{control} = \varnothing$;

\Repeat{$obs\_defs = \varnothing \land \Delta{control} = \varnothing $ }{

     $\mathit{frontiers2check}$=$\mathit{frontiers}(S) - \mathit{checked\_frontiers}$;


     $\mathit{obs\_fr} \gets exercised\_fr\_with\_ctxs(frontiers2Check, x,I^q)$;

     $\mathit{obs\_defs} \gets $ $\{ w \: | \: \exists (w,w_{ctx},r,r_{ctx}) \in \mathit{obs\_fr} \land  r_{ctx} \subseteq S\}$;

     $\mathit{obs\_ctxs} \gets $
     $\{ c \: | \: \exists (w,w_{ctx},r,r_{ctx}) \in \mathit{obs\_fr} \land  r_{ctx} \subseteq S \land c \in w_{ctx}\}$

     $\mathit{kept\_frontiers} \gets $ $\{ (w,r) \: | \: \exists (w,w_{ctx},r,r_{ctx}) \in \mathit{obs\_fr} \land r_{ctx} \not\subseteq S\}$

     $S$ = $S \cup  \mathit{obs\_defs} \cup \mathit{obs\_ctx}$;

     $\Delta{control} = missing\_control(S)$;

     $S = S \cup \Delta{control}$;

    $\mathit{checked\_frontiers}$ =  $\mathit{checked\_frontiers} \cup frontiers(S) - \mathit{kept\_frontiers}$;

}

\KwRet{$S$}\;

\end{algorithm}

\subsubsection{Detecting exercised frontiers}

As mentioned above, the detection  works very similarly to the intraprocedural one with a few extensions. Since the exercised frontiers procedure, in this case, is contextualized 
the instrumentation registers in $M'$ at address $m$ not only the writer node $w$ that is defining $m$ but also the elements of current stack $\pi$. Thus, when a reader of a frontier to check is executed, if the last definition of an used memory address $m$ happens to be a frontier writer node $w$, $M'$ at $m$ also contains the elements of stack snapshot taken at the last $w$ occurrence (the one last defining $m$), as required in the post-condition of this procedure. \vic{Compared to the intraprocedural approach, the memory required is multiplied by the size of a structure for representing sets of calling sites which can be bounded by the maximum number of different call-sites that can make up a call chain in the program. This means, unlike, \textit{execute-once} approaches, worst case memory complexity does not depend on the execution size but linearly on the memory footprint and the number of call sites of the program.}

Regarding libraries, program analyses usually differentiate between client/application code -the code written by the developer- and library code. In particular, libraries pose efficiency and recall challenges to state of the art dynamic slicers \cite{ahmed2021slicer4j}.
Our approach can be naturally adapted to avoid analyzing frontiers inside library code. In essence, as \cite{ahmed2021slicer4j} we leverage library models of static taint analyzers, usually called summaries (e.g.,\cite{arzt2016stubdroid,sridharan2011f4f, codeql}, etc.). In our case, those models are used by the $SDA$ component to infer data dependencies at client code level without analyzing library code. The actual adaptation of the approach is to use the summaries as actual read and write operations by wrapping the calls to libraries methods, similarly to \cite{ahmed2021slicer4j}. This is done as an adaptation of our frontier detection instrumentation. It is worth noting that this manipulation is sound provided that the summary's output indicates a ``MUST define'' behavior, as our detection instrumentation would assume a redefinition when updating last def information.

\section{Evaluation}
The first research question we aim to address is:

\begin{itemize}
    \item [\textbf{RQ1}] How does our approach compare with \textit{execute-once} dynamic slicers with respect to performance, precision and recall?
\end{itemize}

Our second research question aims at exploring the extent to which our approach depends more on slice size and less on execution size than \textit{execute-once} approaches. Precision and recall cannot be affected by execution and slice size, thus we focus on performance.


\begin{itemize}
     \item [\textbf{RQ2}] How does execution  and slice size impact the performance gains and losses of our approach with respect to \textit{execute-once} dynamic slicers?
\end{itemize}

\tse{
Our approach is different to \textit{execute-once} in two ways. On one hand we corroborate dependencies, rather than register and traverse explicit memory locations. On the other, we do corroboration \textit{on-demand} by re-executions. Our third question aims to explore to what extent each contribute to shifting the dependency of our approach towards slice size. To isolate the impact in performance of dependency corroboration, we compare our \textit{on-demand} approach with one in which all potential dependencies are corroborated \textit{upfront} in one single execution.}


\begin{itemize}
     \item [\textbf{RQ3}] How does \textit{on-demand} dependency corroboration compare to \textit{upfront} dependency corroboration with respect to performance?
\end{itemize}

To evaluate our approach, we developed an implementation of the on-demand re-execution inter-procedural algorithm that targets the Java programming language. Our implementation uses a static dependency analysis built on the publicly available static analysis tool CodeQL \cite{codeql} for frontier detection. This analyzer is to the best of our knowledge a flow-sensitive, object-insensitive, context-insensitive static taint analyzer implemented on Datalog that compromises precision to achieve scale. Our frontier tracker is built using a custom Java-agent that uses ASM \cite{asm} for runtime code instrumentation, and Java IdentityHashMaps and JVMTI tagging technology~\cite{jvmti} for low level def-use coverage checking.

We compare our implementation against Javaslicer~\cite{javaslicer}, a state-of-the-art dynamic slicer for Java. Javaslicer implements the algorithm described by Wang and Roychoudhury \cite{wang2004using,wang2008dynamic}, that can be thought of as a Korel-Laski implementation that works at bytecode-level and that slices by performing a closure of data and control dependencies but skips the Korel-Laski identity relation. Javaslicer usage is as follows: $i)$ Tracing: the target Java program must be run once with the Javaslicer instrumentation agent, that on termination outputs a trace file where it stores every memory references that was accessed (read or write) by each executed bytecode instruction. $ii)$ Slicing: Once a trace is produced, Javaslicer provides an executable jar that given a trace file and a node criteria (a line number, and optionally a set of variables) slices the trace backwards and outputs the program slice as a list of bytecode instructions.

We also compare against Slicer4J that is a recently published slicer whose main goal is to support the latest Java features that Javaslicer does not support out-of-the-box. Slicer4j authors acknowledge that for this they pay a price in terms of performance in comparison with Javaslicer~\cite{ahmed2021slicer4j}.

We do not compare against DrDebug as it only supports C/C++ making comparison against Javaslicer, Slicer4J and our implementation very challenging. Some experiments we did provided some indication DrDebug would not be significantly different.

\tse{
Specifically for RQ3, we compare our on-demand slicer against two components that performs \textit{upfront} dependency corroboration: \textit{upfront-all}, and  \textit{upfront-slice}. We developed these two components with the same dynamic instrumentation code than our approach and using the same static dependency analysis (to avoid any coding bias), but targeting all potentially relevant dependencies upfront. Concretely, these components differ by the set of dependencies targeted for corroboration: The \textit{upfront-all} component corroborates all the static data dependencies found by the static dependency analysis, and the \textit{upfront-slice} component focuses on the data dependencies that are present in a static slice of the criterion. Note that these components do not include the logic to yield slices like the on-demand tool and are meant to isolate the corroboration logic. Thus, being a sub-task of an hypothetical slicing based on those corroboration approaches, they are fair proxies to analyze our speed ups and degree of coupling with respect to execution and slice size.
}


We use two benchmarks for the evaluation. The first is SV-COMP benchmark from which we took all its Java algorithms. These are short but computationally intensive programs that manipulate data structures (e.g., insertion sort, bellman-ford, red-black-tree, binary-tree, etc.). The benchmark is suitable to perform a controlled evaluation for various reasons: $i)$ The algorithms have simple inputs (e.g., a list of integers) which straightforwardly impact the execution size, and $ii)$ being short programs, it is feasible to cover all possible slice criteria to avoid bias in criteria selection while varying slice size.



The choice of the second benchmark was motivated by using a large real software application with slicing criteria derived from real dynamic slicing use cases~\cite{yoo2012regression}. We selected Antlr4~\cite{antlr4}, as it is Github's top starred project where Javaslicer has already been applied~\cite{javaslicer}, and has a test suite that includes assertions that can be used as proxies for some sort of real slices that a user may require for debugging, regression testing or program understanding. Antlr4 is a parser generator for reading, processing, executing, or translating structured text or files. It is widely used to build languages, tools, and frameworks. From a grammar, Antlr4 generates a parser that can build and walk parse trees.

For Antlr4 we used as slicing criteria the first 50 assertions that are executed in its suite of unit-tests. Unit-test tend to favour small execution sizes in which code that is irrelevant to the test purpose (i.e., the assertion) is not best practice. In in other words, these tests will tend to minimize the difference between slice sizes and execution size, which in principle is biased towards execute-once approaches. We also included Antlr4's only performance test, which is of the opposite nature. This test included one assertion that was used as slicing criteria.
\tse{Due to the very different nature of the Antlr4 tests, we report results for Antlr4 unit tests and performance tests separately.}

To measure efficiency we used the resource python API~\cite{pythonresource}, executing processes sequentially and averaging runs. We also measured the average disk space used by Javaslicer for trace storage.
We ran all our experiments on a computer with a i7-856U CPU, 500GB solid state drive and 16GB of RAM.
\subsection{SV-COMP}

To perform a thorough and fair comparison between our approach, Javaslicer and Slicer4J we replaced $\mathtt{nondet()}$ input reads with command line inputs to define a fixed data set. We fixed the maximum problem sizes so to avoid memory thrashing of Javaslicer which downgrades its performance considerably (recall that Javaslicer is reported to perform better than Slicer4J~\cite{ahmed2021slicer4j}). Problem specific arguments were fixed randomly.


To select slicing criteria for each program we executed the first input of its data set and collected every executed line as a criterion. Thus, for every subject we computed a different number of slices. In total we built 1784 different slices using our approach and Javaslicer.

\begin{table*}[]
\caption{Evaluation results.}

\resizebox{\textwidth}{!}{%

\begin{tabular}{lcccccccccccccccc}
\hline
\multicolumn{1}{|c|}{\multirow{3}{*}{\textbf{Subject}}} & \multicolumn{1}{c|}{\multirow{3}{*}{\begin{tabular}[c]{@{}c@{}}N°\\ of\\ Slices\end{tabular}}} & \multicolumn{1}{c|}{\multirow{3}{*}{\begin{tabular}[c]{@{}c@{}}Exec. \\ Time(s)\end{tabular}}} & \multicolumn{1}{c|}{\multirow{3}{*}{\begin{tabular}[c]{@{}c@{}}Javaslicer \\ Tracing \\   Time(s)\end{tabular}}} & \multicolumn{1}{c|}{\multirow{3}{*}{\begin{tabular}[c]{@{}c@{}}Javaslicer \\ Slicing \\ Time(s)\end{tabular}}} & \multicolumn{1}{c|}{\multirow{3}{*}{\begin{tabular}[c]{@{}c@{}}Slice \\ Size \\ (Loc)\end{tabular}}} & \multicolumn{1}{c|}{\multirow{3}{*}{\begin{tabular}[c]{@{}c@{}}On-De \\ Re-ex\\ Time(s)\end{tabular}}} & \multicolumn{1}{c|}{\multirow{3}{*}{\begin{tabular}[c]{@{}c@{}}Gains \\ $\frac{C+D}{F}$\end{tabular}}} & \multicolumn{1}{c|}{\multirow{3}{*}{\begin{tabular}[c]{@{}c@{}}Recall \\ Loss\end{tabular}}} & \multicolumn{1}{c|}{\multirow{3}{*}{\begin{tabular}[c]{@{}c@{}}Prec. \\ Loss\end{tabular}}} & \multicolumn{1}{c|}{\multirow{3}{*}{\begin{tabular}[c]{@{}c@{}}Trace \\ Size \\ (MB)\end{tabular}}} & \multicolumn{1}{c|}{\multirow{3}{*}{\begin{tabular}[c]{@{}c@{}}N° \\  of \\ Re-ex\end{tabular}}} & \multicolumn{1}{c|}{\multirow{3}{*}{\begin{tabular}[c]{@{}c@{}}N° \\  of \\ $f2c$\end{tabular}}} & \multicolumn{2}{c|}{\begin{tabular}[c]{@{}c@{}}upfront\\ $\mathit{all}$\end{tabular}} & \multicolumn{2}{c|}{\begin{tabular}[c]{@{}c@{}}upfront\\ $\mathit{slice}$\end{tabular}} \\ \cline{14-17}
\multicolumn{1}{|c|}{} & \multicolumn{1}{c|}{} & \multicolumn{1}{c|}{} & \multicolumn{1}{c|}{} & \multicolumn{1}{c|}{} & \multicolumn{1}{c|}{} & \multicolumn{1}{c|}{} & \multicolumn{1}{c|}{} & \multicolumn{1}{c|}{} & \multicolumn{1}{c|}{} & \multicolumn{1}{c|}{} & \multicolumn{1}{c|}{} & \multicolumn{1}{c|}{} & \multicolumn{1}{c|}{\multirow{2}{*}{Time(s)}} & \multicolumn{1}{c|}{\multirow{2}{*}{Deps}} & \multicolumn{1}{c|}{\multirow{2}{*}{Time(s)}} & \multicolumn{1}{c|}{\multirow{2}{*}{Deps}} \\
\multicolumn{1}{|c|}{} & \multicolumn{1}{c|}{} & \multicolumn{1}{c|}{} & \multicolumn{1}{c|}{} & \multicolumn{1}{c|}{} & \multicolumn{1}{c|}{} & \multicolumn{1}{c|}{} & \multicolumn{1}{c|}{} & \multicolumn{1}{c|}{} & \multicolumn{1}{c|}{} & \multicolumn{1}{c|}{} & \multicolumn{1}{c|}{} & \multicolumn{1}{c|}{} & \multicolumn{1}{c|}{} & \multicolumn{1}{c|}{} & \multicolumn{1}{c|}{} & \multicolumn{1}{c|}{} \\ \hline
\multicolumn{1}{c}{} & A & B & C & D & E & F & G & H & I & J & K & L & M & N & O & P \\ \hline
\multicolumn{1}{|l|}{MergeSort} & \multicolumn{1}{c|}{120} & \multicolumn{1}{c|}{0.063} & \multicolumn{1}{c|}{33.21} & \multicolumn{1}{c|}{84.73} & \multicolumn{1}{c|}{9.15} & \multicolumn{1}{c|}{1.80} & \multicolumn{1}{c|}{114.64} & \multicolumn{1}{c|}{0.07} & \multicolumn{1}{c|}{0.03} & \multicolumn{1}{c|}{59.13} & \multicolumn{1}{c|}{2.83} & \multicolumn{1}{c|}{4.35} & \multicolumn{1}{c|}{6.9} & \multicolumn{1}{c|}{36} & \multicolumn{1}{c|}{2.24} & \multicolumn{1}{c|}{9.5} \\ \hline
\multicolumn{1}{|l|}{BellmanFord} & \multicolumn{1}{c|}{120} & \multicolumn{1}{c|}{0.057} & \multicolumn{1}{c|}{45.36} & \multicolumn{1}{c|}{18.49} & \multicolumn{1}{c|}{15.07} & \multicolumn{1}{c|}{1.47} & \multicolumn{1}{c|}{51.08} & \multicolumn{1}{c|}{0.14} & \multicolumn{1}{c|}{0.13} & \multicolumn{1}{c|}{3.50} & \multicolumn{1}{c|}{3.77} & \multicolumn{1}{c|}{3.80} & \multicolumn{1}{c|}{1.3} & \multicolumn{1}{c|}{26} & \multicolumn{1}{c|}{0.48} & \multicolumn{1}{c|}{8.47} \\ \hline
\multicolumn{1}{|l|}{RedBlackTree} & \multicolumn{1}{c|}{907} & \multicolumn{1}{c|}{0.092} & \multicolumn{1}{c|}{52.26} & \multicolumn{1}{c|}{34.34} & \multicolumn{1}{c|}{63.31} & \multicolumn{1}{c|}{12.66} & \multicolumn{1}{c|}{17.28} & \multicolumn{1}{c|}{0.17} & \multicolumn{1}{c|}{0.12} & \multicolumn{1}{c|}{17.55} & \multicolumn{1}{c|}{6.41} & \multicolumn{1}{c|}{91.3} & \multicolumn{1}{c|}{3.8} & \multicolumn{1}{c|}{218} & \multicolumn{1}{c|}{0.74} & \multicolumn{1}{c|}{89} \\ \hline
\multicolumn{1}{|l|}{BinaryTree} & \multicolumn{1}{c|}{324} & \multicolumn{1}{c|}{0.044} & \multicolumn{1}{c|}{38.78} & \multicolumn{1}{c|}{10.35} & \multicolumn{1}{c|}{16.66} & \multicolumn{1}{c|}{0.98} & \multicolumn{1}{c|}{64.60} & \multicolumn{1}{c|}{0.13} & \multicolumn{1}{c|}{0.02} & \multicolumn{1}{c|}{2.71} & \multicolumn{1}{c|}{2.94} & \multicolumn{1}{c|}{16.1} & \multicolumn{1}{c|}{0.7} & \multicolumn{1}{c|}{26} & \multicolumn{1}{c|}{0.5} & \multicolumn{1}{c|}{18} \\ \hline
\multicolumn{1}{|l|}{InsertionSort} & \multicolumn{1}{c|}{66} & \multicolumn{1}{c|}{0.068} & \multicolumn{1}{c|}{60.40} & \multicolumn{1}{c|}{206.55} & \multicolumn{1}{c|}{6.71} & \multicolumn{1}{c|}{5.01} & \multicolumn{1}{c|}{168.78} & \multicolumn{1}{c|}{0.10} & \multicolumn{1}{c|}{0.00} & \multicolumn{1}{c|}{271.00} & \multicolumn{1}{c|}{2.69} & \multicolumn{1}{c|}{4.09} & \multicolumn{1}{c|}{42.1} & \multicolumn{1}{c|}{11} & \multicolumn{1}{c|}{16.1} & \multicolumn{1}{c|}{5.38} \\ \hline
\multicolumn{1}{|l|}{SortedList} & \multicolumn{1}{c|}{127} & \multicolumn{1}{c|}{0.055} & \multicolumn{1}{c|}{56.83} & \multicolumn{1}{c|}{26.90} & \multicolumn{1}{c|}{4.57} & \multicolumn{1}{c|}{0.51} & \multicolumn{1}{c|}{236.98} & \multicolumn{1}{c|}{0.00} & \multicolumn{1}{c|}{0.00} & \multicolumn{1}{c|}{6.83} & \multicolumn{1}{c|}{2.43} & \multicolumn{1}{c|}{3.02} & \multicolumn{1}{c|}{1.6} & \multicolumn{1}{c|}{13} & \multicolumn{1}{c|}{0.8} & \multicolumn{1}{c|}{7.5} \\ \hline
\multicolumn{1}{|l|}{Tsp} & \multicolumn{1}{c|}{120} & \multicolumn{1}{c|}{0.084} & \multicolumn{1}{c|}{45.41} & \multicolumn{1}{c|}{251.29} & \multicolumn{1}{c|}{13.91} & \multicolumn{1}{c|}{12.08} & \multicolumn{1}{c|}{217.81} & \multicolumn{1}{c|}{0.04} & \multicolumn{1}{c|}{0.05} & \multicolumn{1}{c|}{355.8} & \multicolumn{1}{c|}{3.42} & \multicolumn{1}{c|}{8.00} & \multicolumn{1}{c|}{58.9} & \multicolumn{1}{c|}{28} & \multicolumn{1}{c|}{23.13} & \multicolumn{1}{c|}{14.08} \\ \hline
\multicolumn{1}{|l|}{Total \& Avg.} & \multicolumn{1}{c|}{1784} & \multicolumn{1}{c|}{0.066} & \multicolumn{1}{c|}{47.46} & \multicolumn{1}{c|}{90.38} & \multicolumn{1}{c|}{18.48} & \multicolumn{1}{c|}{4.93} & \multicolumn{1}{c|}{124.45} & \multicolumn{1}{c|}{0.09} & \multicolumn{1}{c|}{0.05} & \multicolumn{1}{c|}{102.36} & \multicolumn{1}{c|}{3.51} & \multicolumn{1}{c|}{18.6} & \multicolumn{1}{c|}{16.5} & \multicolumn{1}{c|}{51} & \multicolumn{1}{c|}{6.28} & \multicolumn{1}{c|}{21.7} \\ \hline \hline
\multicolumn{1}{|l|}{Antlr4 Unit} & \multicolumn{1}{c|}{50} & \multicolumn{1}{c|}{0.01} & \multicolumn{1}{c|}{51.28} & \multicolumn{1}{c|}{34.56} & \multicolumn{1}{c|}{57.75} & \multicolumn{1}{c|}{10.75} & \multicolumn{1}{c|}{8.23} & \multicolumn{1}{c|}{0.04} & \multicolumn{1}{c|}{0.7} & \multicolumn{1}{c|}{16.07} & \multicolumn{1}{c|}{11.08} & \multicolumn{1}{c|}{427.7} & \multicolumn{1}{c|}{43.09} & \multicolumn{1}{c|}{88K} & \multicolumn{1}{c|}{41.5} & \multicolumn{1}{c|}{66K} \\ \hline
\multicolumn{1}{|l|}{Antlr4 Perf.} & \multicolumn{1}{c|}{1} & \multicolumn{1}{c|}{1.78} & \multicolumn{1}{c|}{120} & \multicolumn{1}{c|}{T/O} & \multicolumn{1}{c|}{54} & \multicolumn{1}{c|}{5m} & \multicolumn{1}{c|}{N/A} & \multicolumn{1}{c|}{N/A} & \multicolumn{1}{c|}{N/A} & \multicolumn{1}{c|}{3GB} & \multicolumn{1}{c|}{7} & \multicolumn{1}{c|}{53} & \multicolumn{1}{c|}{T/O} & \multicolumn{1}{c|}{-} & \multicolumn{1}{c|}{T/O} & \multicolumn{1}{c|}{-} \\ \hline
\end{tabular}
}

\label{tabla}
\end{table*}

\begin{figure*}[htbp]
  \begin{minipage}[t]{8cm}
  \centering
    \includegraphics[scale=.30]{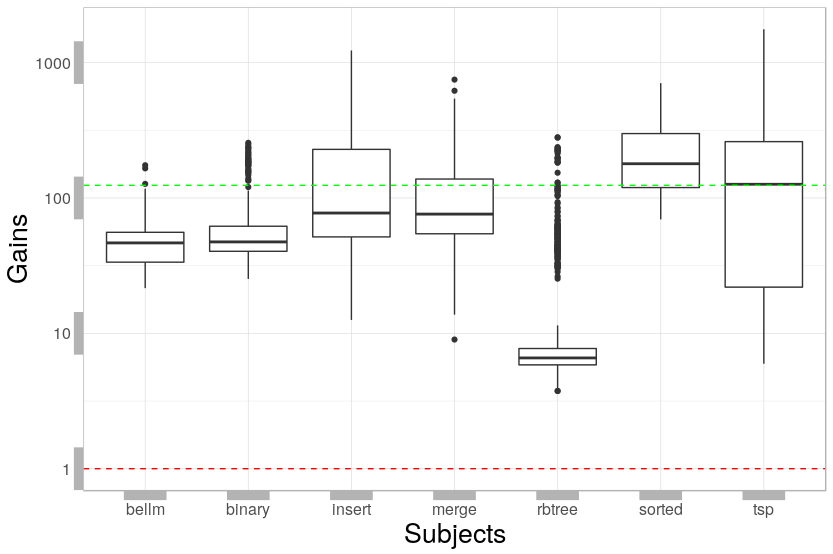}
    \caption{Performance gains in SV-COMP subjects. Axis $y$ is in logarithmic scale.}
    \label{box-plot-svcomp}
    \end{minipage}
    \hspace{0.5cm}
      \begin{minipage}[t]{8cm}
      \centering
    \includegraphics[scale=.30]{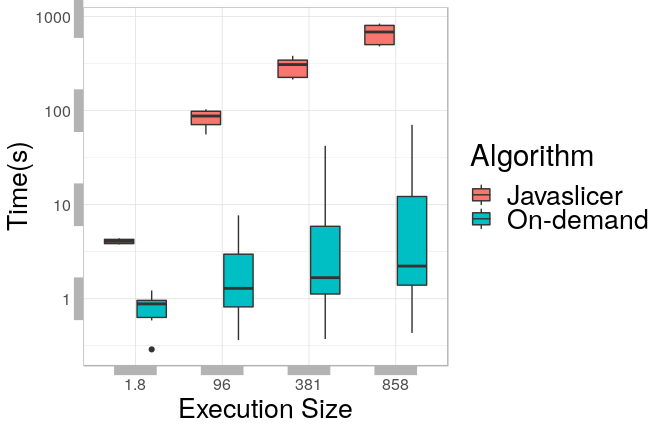}
    \caption{Performance (in seconds) for InsertionSort  grouping slices by execution size (file size in Mb). Axis $y$ is in logarithmic scale.}
     \label{insertion-box-plot}
    \end{minipage}\\
      \begin{minipage}[t]{8cm}
  \centering
    \includegraphics[scale=.35]{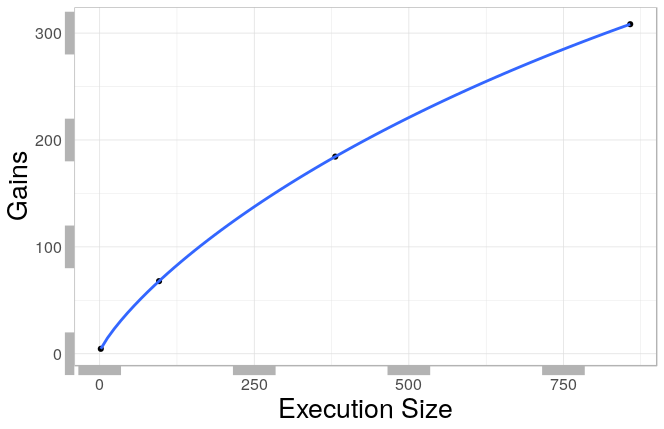}
    \caption{Median performance gain for slices grouped by execution size (file size in Mb) for InsertionSort, fitted with a linear model of $x * log(x)$ curve with $R^2$=0.99.}
    \label{median-gain}
    \end{minipage}
    \hspace{0.5cm}
      \begin{minipage}[t]{8cm}
      \centering
    \includegraphics[scale=.30]{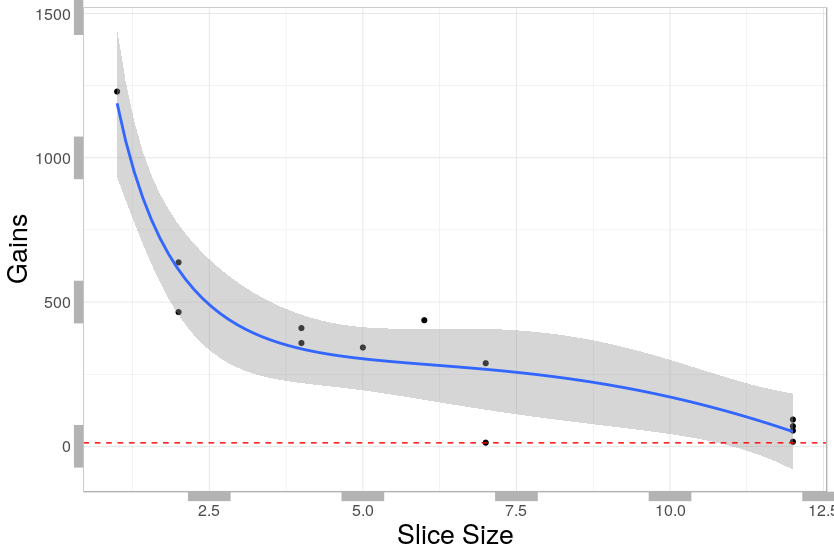}
    \caption{InsertionSort performance gain by slice size when processing its biggest trace size (858Mb), fitted with a linear model of $x * log(x)$ curve with $R^2$=0.87. }
     \label{merge-sort-slice-size}
    \end{minipage}\\
          \begin{minipage}[t]{8cm}
  \centering
    \includegraphics[scale=.30]{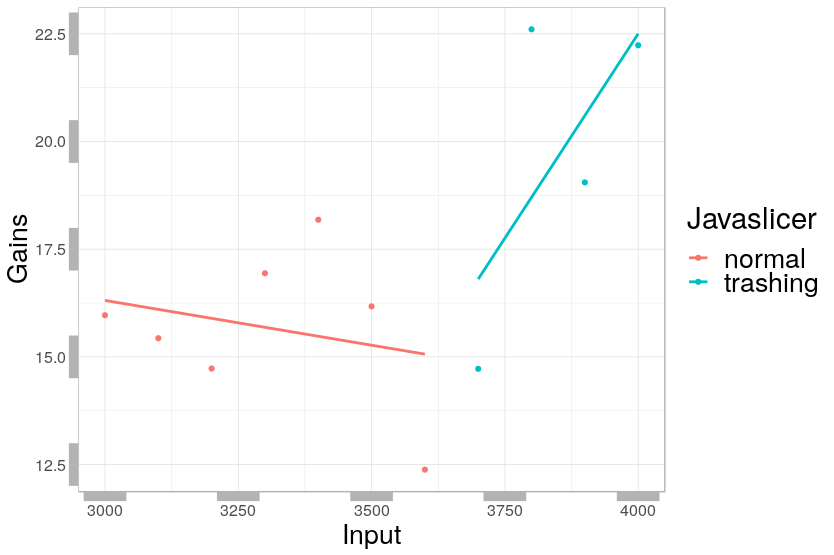}
    \caption{InsertionSort performance gains when slicing the biggest subject slice (in LoC) with inputs between 3000 and 4000 elements. Green/orange are executions for which Javaslicer does/does-not use swap memory, fitted with two different linear regressions.}
    \label{worst-case}
    \end{minipage}
    \hspace{0.5cm}
      \begin{minipage}[t]{8cm}
    \includegraphics[scale=.30]{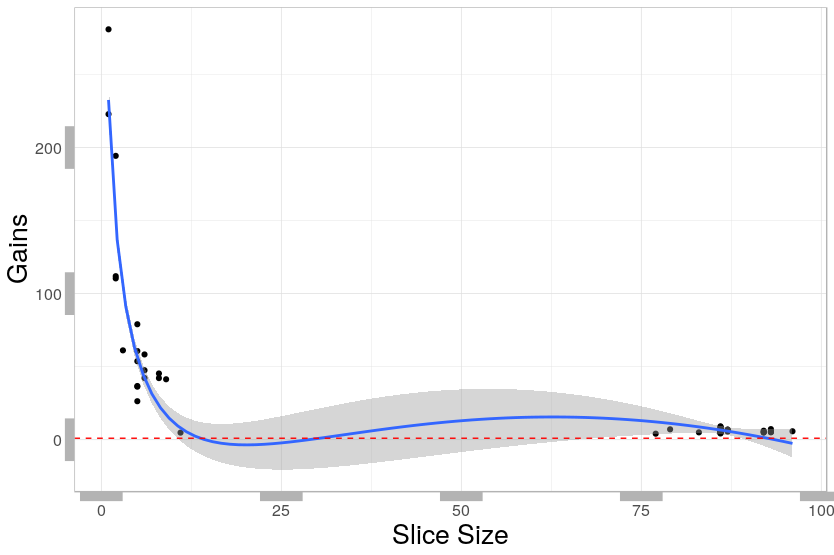}
    \caption{RedBlackTree performance gain vs.  slice size, when processing its biggest trace size, fitted with a linear model of $x * log(x)$ curve with $R^2$=0.92. }
     \label{red-black-tree-slice-size}
    \end{minipage}

\end{figure*}

\subsubsection{RQ1}

In Table~\ref{tabla} we report results of slice construction for the SV-COMP benchmark for Javaslicer and our implementation. We report for each subject collected data in following order: the number of slices we built (col. A), the average baseline execution time for each input (col. B), the time that Javaslicer took on average for tracing each input (col. C), the average time that Javaslicer spent computing a slice based on traced information (col. D), the average slice size (lines of code) computed by Javaslicer (col. E), the average time per slice spent by our approach (col. F), the average of the gains per slice when compared with Javaslicer (tracing plus slicing) (col. G), the average number of nodes per slice that Javaslicer includes but that our approach does not --referred to as recall loss-- (col. H), the average number of nodes per slice that our approach includes in the slice but that Javaslicer does not --referred to as precision loss-- (col. I), and the average Javaslicer trace file size (col. J). \tse{Each MB of a Javaslicer trace corresponds to approximately 0.4M executed instructions.}


The table shows a performance gain of our approach over all subjects (avg 124x, max $>$1000x, min 3.74x), with reasonably low recall loss (avg ~9\%, max 17\%, min 0\%) and low precision loss (avg 5\%, max 13\%, min 0\%), suggesting that our approach could be more efficient than the current state-of-art in dynamic slicing at least in small stand alone subjects. On-demand re-execution seems to be a perfectly valid penalty in terms of efficiency compared with traditional trace storing and traversing.

Recall loss is due to how Javaslicer computes control dependencies for loops \cite{ranganath2005new}. Every loop is considered by Javaslicer as a potential non terminating cycle in the control flow graph, and thus controls whenever any successor node is executed or not. However, in practice this definition always adds all executed loops in the slice, even if their execution is irrelevant to the slice criteria computation.
Given that both Javaslicer and our approach are designed to slice terminating executions, this control dependency adds nothing relevant to the slice and can be understood as an imprecision loss Javaslicer rather than a recall loss of our approach. Moreover, this imprecision is amplified as loop variables and loop conditions depend transitively on other nodes. Consistent with this observation is that the only subject with 0\% recall loss, SortedList, has no successors after its only loop.

Regarding precision loss our findings show that Javaslicer, being designed to slice Java bytecode, chooses to model each method call argument as a separated node regardless of being in a single line at the source-code level. This causes Javaslicer for a method that has $n$-parameters to add the method call and only the parameters actually used in the execution. However at source-code granularity every parameter has to be in the slice. \tse{For instance, one way of obtaining the exact same results as Javaslicer would be to transform all parameters of the program's methods to global variables and assign them in different lines before each call-site}.

We also looked into a multiple slicing criteria scenario, where Javaslicer can take advantage of tracing once and then traverse the same trace for each criterion. For each input, we computed one trace using Javaslicer and then built a slice for every criterion from that trace (avg. 32 criteria per trace). On average our implementation, computing slices for all criteria, exhibited a 67x speed up.

In Figure~\ref{box-plot-svcomp}, we depict a box-plot that shows in more detail how performance gain varies for the different SV-COMP algorithms. These variations are the concern of RQ2.

We do not include detailed performance results for Slicer4J as these were significantly worse that Javaslicer and our implementation. For over 70\% of slice criteria Slicer4J timed-out. For criteria that was successfully sliced (those with the smallest execution size), computation time was on average 50x (max 418x, min 8x) compared to our implementation.

\subsubsection{RQ2.}

We discuss variations in efficiency of InsertionSort slices with respect to execution and slice size as these are representative of all the rest except Red Black Tree. We then discuss Red Black Tree to explain why it has a significantly different box plot in Figure~\ref{box-plot-svcomp}.

Figure~\ref{median-gain} shows how the median gain for InsertionSort increases with execution size. However, this can be slightly misleading as the execution time dispersion of our approach grows with execution size (see Figure~\ref{insertion-box-plot}).

\tse{
We conducted a follow-up analysis on InsertionSort to study our gains with respect to execution size: we grouped slices by slicing criteria and we ranked them in each group by execution size and by overall gains to compute the Spearman correlation coefficient. We observed a positive correlation coefficient for 75\% of the groups (the bigger the execution size the bigger the gain of our approach) with statistical significance ($p < 0.005$), 12.5\% do not exhibit correlation (gains remain constant, the execution size affects our approach the same as Javaslicer), and for the remaining 12.5\% we observe a negative correlation (gains drop while execution size increases). The latter correspond to cases where the slice size is the complete code, which is our worst case. Indeed, Figure~\ref{merge-sort-slice-size} shows that worst gains are achieved with slices that include all (i.e., 12) executed statements.
}

We conducted an additional experiment for InsertionSort picking these worst cases and augmenting the execution size (i.e., parameters that force loops to be executed more times) to stress test our approach.  Figure~\ref{worst-case} shows that as execution size increases gains are reduced (but are always positive) up to the point in which Javaslicer occupies its entire RAM memory allocation and starts swapping.

The reason why Red Black Tree has a notably different box-plot in Figure~\ref{box-plot-svcomp} is due to the fact that all slices involved similar execution sizes thus efficiency gain variations may be mainly impacted by slice size which Figure~\ref{red-black-tree-slice-size} shows to be bi-modal.

We conducted another follow-up analysis: for each subject we grouped the slices with same input (thus same execution size) and we ranked them in each group by slice size and by overall gains to compute the Spearman correlation coefficient. For every group we observed a negative correlation coefficient (the smaller the slice the bigger the gain) with strong statistical significance ($p < 0.001$).

Regarding the number of re-executions, we observe that on average, they are sub-linear with respect to the slice size. This is expected as our approach tracks multiple frontiers in each re-execution, which leads to adding multiple nodes at once in each iteration.

\begin{figure}
    \centering
    \includegraphics[scale=0.35]{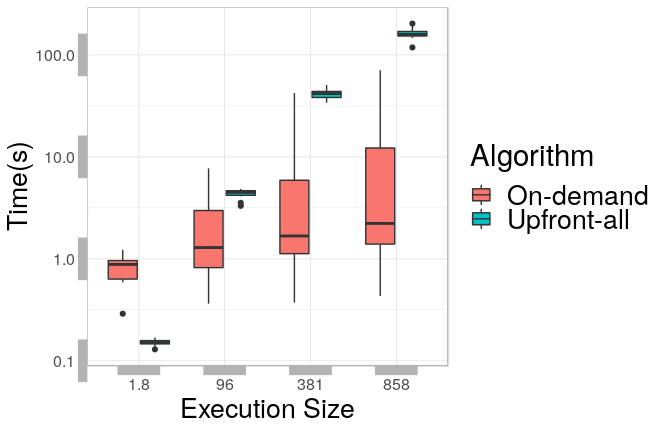}
    \caption{InsertionSort on-demand re-execution slicing approach performance vs. upfront-all dependency corroboration. Axis $y$ is in logarithmic scale.}
    \label{upfront-insertion-sort}
\end{figure}

\subsubsection{RQ3.}
In Table~\ref{tabla}, we report a comparison between our approach and the upfront dependency corroboration programs. We report: the average number of re-executions per slice (col. K), the average number of on-demand frontiers to corroborate per slice (col. L), the average execution time of the upfront dependency corroboration programs per slice (col. M and col. O), and the average number of dependencies to corroborate upfront per slice (col. N and col. P).
The results show an overall performance gain between our approach and the upfront corroborations programs (avg 3.34x with \textit{upfront-all}, and avg 1.25x with \textit{upfront-slice}). We conducted a follow-up analysis to study these gains and their dependency with execution size and slice size. We found that \textit{upfront-all} performance is largely dependent on execution size, see Figure~\ref{upfront-insertion-sort}, reporting a similar pattern than Javaslicer (the bigger the execution size, the bigger the gain of our approach). On the other hand, \textit{upfront-slice} performance depends less on execution size. In particular, in cases where the static slice is small (i.e, precise), \textit{upfront-slice} reports similar or slightly better times than our approach. However, \textit{upfront-slice} reports a stronger dependency on execution size than our approach when the static slice is large and imprecise. In these cases, the static slice removes few dependencies and the performance is similar to that of \textit{upfront-all}. This provides some indication that on-demand corroboration can improve performance when static slicing has precision problems.
\subsection{Antlr4 Unit Tests}
\label{antlr-eval}


\begin{figure}
    \centering
    \includegraphics[scale=0.30]{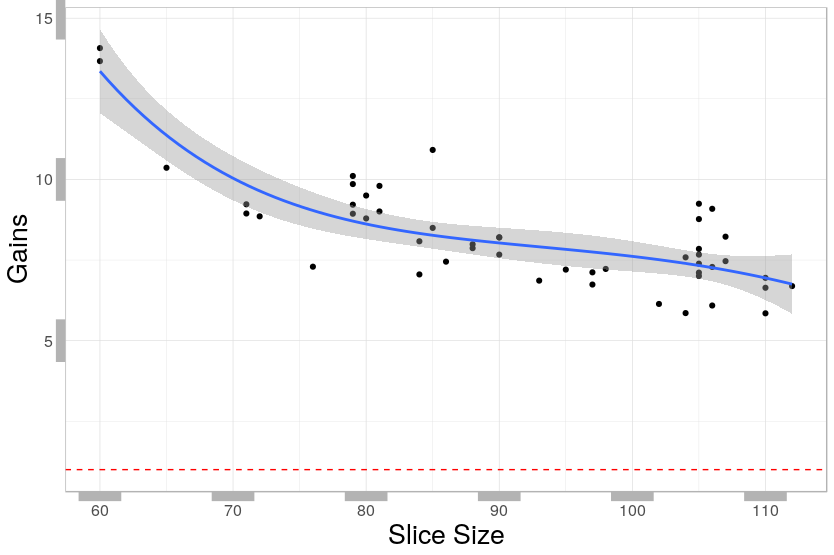}
    \caption{Antlr4 performance gain vs. slice size, when processing unit tests, fitted with a linear model of $x * log(x)$ curve with $R^2$=0.66.}
    \label{antlr-gains-per-slice-size}
\end{figure}

As Antlr4 has external libraries, our approach requires providing summaries for them. Summaries for some classes from the Java Standard Library (e.g., lists, maps, etc.) were taken out-of-the-box from the distribution of CodeQL, others were developed manually.



\subsubsection{RQ1.}
We sliced the first 50 test assertions of the Antlr4 test suite. The average slice size output by Javaslicer was 57.75.
On the other hand, Slicer4J crashed at every test case.

Regarding efficiency, we found that the average gain was 8.24, the standard deviation was 1.68, the minimum 5.84 and maximum 14. While the total running time for Javaslicer was 70 minutes, it was 9 minutes for our approach.  On average, our approach performed 11 re-executions per slice.
Regarding recall, we observed an average 4\% loss which we attribute to $i)$ lack of modelling of some libraries and $ii)$ differences in control dependencies similarly to the ones reported in the SV-COMP experiment. Regarding precision, the reported loss is on average 70\%. However, we consider this difference to be all due to accidental differences as with the SV-COMP benchmark: We explored manually the five slices for which our approach showed maximum precision loss. We found that the cause for the loss was the same as in the SV-COMP benchmark regarding differences in how a method's call arguments are modelled and sliced. 

It is important to note the average gain (8x) for the Antlr4 unit tests is lower than the worst average gain amongst the SV-COMP programs, namely RedBlackTree (21x). An explanation for this lies in understanding how execution and slice size impacts slicing the Antlr4 unit tests (i.e., RQ2).

As with SV-COMP, we also analysed a multiple-criteria scenario. In Antlr4, each test on average had 4 criteria, which we sliced together for each input. Our approach showed a speedup of 5x compared to Javaslicer.

\subsubsection{RQ2.}
To better address this research question we performed the same follow up analysis used for SV-COMP RQ2. We ranked the tests according to slice sizes and gain ratio,
obtaining a Spearman correlation coefficient of -0.65 over an n=50 which has a strong statistical significance of $p < 0.001$. Therefore replicating the same correlated behaviour that we observed with the SV-COMP benchmark: Smaller sized slice achieve better gains as showed in Figure~\ref{antlr-gains-per-slice-size}.

Unfortunately, the relation between execution size and gains cannot be studied with these 50 unit test slices because their execution size differs insignificantly. We opt for not manually changing the tests to impact execution size to keep the treatment of Antlr4 slices unbiased. For this very reason, we also selected the only performance test of Antrl4.

\subsubsection{RQ3}
We compared our approach with \textit{upfront-all} and \textit{upfront-slice} corroboration programs. Results showed that our approach is at least 4x faster to produce a slice when compared with upfront dependency corroboration. Interestingly, \textit{upfront-all} and \textit{upfront-slice} report similar times, this happens because they target the same dependencies for corroboration due to static slice imprecision ($88$K dependencies and $66$K respectively). Both upfront programs use the same library summaries that we use for our approach to avoid any bias related to libraries.


\subsection{Antlr4 Performance Test}
\subsubsection{RQ1.}
We sliced the Antlr4 performance test with Javaslicer and our approach. The execution of the test with no instrumentation takes 20 seconds. Results regarding RQ1 are that Javaslicer produced a 3GB trace file (consistent with a large execution size) and after 15 minutes of traversing the trace it ran out of memory (16GB). On the other hand, our approach sliced the test assertions in 5 minutes on average, re-executing the program 8 times on average. The fact that our approach scales to these slicing tasks while Javaslicer does not is consistent with the observation that in general our approach obtains increased gain with bigger executions.

\subsubsection{RQ2 and RQ3}
With only one slicing criteria for the only performance test, there is insufficient data to investigate the correlation between execution or slice size and gains.
\tse{Both upfront-all and upfront-slice programs timed out after 15 minutes.}


\subsection{Summary of Results}


Regarding RQ1, our evaluation provides evidence that our implementation of the on-demand re-execution paradigm for dynamic slicing results in performance gains with respect to \textit{execute-once} tools with comparable precision and recall.

For the 7 algorithms implemented in Java of the SV-COMP benchmark, with more than 1700 unique combinations of inputs and target criteria, we show a significant speedup (avg 124x, max $>$1000x, min 3.74x)  when compared with Javaslicer. We also show gains for multi-criteria usage scenarios. Slicer4J performed significantly worse than both approaches.

For the Antlr4 case study, we report results for the construction of 50 slices from real unit test assertions where we obtain, when compared to Javaslicer, an average 8x gain in overall slice time. We also report gains in multi-criteria usage scenarios. Slicer4J failed to slice Antlr4 unit tests. We also report the slicing of Anltr4's performance test, in this case Javaslicer ran out of memory while our approach sliced the test assertion in 5m.

For RQ2, results provide positive evidence that the gains of our implementation of on-demand re-execution compared to execute-once approaches is directly proportional to execution size and inversely proportional to slice size. This is observable in all benchmarks. In particular, for SV-COMP and Antlr4 unit tests we observe for comparable execution sizes, a negative Spearman correlation between slice size and performance gains with a statistical significance of $p < 0.001$. While for comparable slices, with the same criterion, in SV-COMP we observe that our approach reports a positive correlation between execution size and gains for 75\% of the slices with a statistical significance of $p < 0.005$. In Antlr4, we observe that execute-once tools have scale problems slicing an execution of a performance test.

Regarding RQ3, results provide evidence that on-demand corroboration offers gains compared to upfront dependency corroborations. Results indicate that gains increase with execution size and imprecision of dependencies to be corroborated up-front. Indeed, in Antlr4 unit tests we observe that nearly half of the gains associated to our approach are due to on-demand corroboration, as upfront-all and upfront-slice are both 4x slower. While at Antlr4 performance test, neither upfront component scale. This provides evidence that on-demand dependency corroboration plays an important role in our approach performance and its shift in practical complexity to slice size rather than execution size.

\section{Threats to Validity}
To slice Antlr4 our approach required libraries summaries, some of which were developed manually others by a third party, which may lead to diminished recall (but experimentation shows this small and accountable to differences with Javaslicer), to increased frontier dependency tracking (which is detrimental to the performance of our approach). The differences in precision and recall between our approach and Javaslicer were explained by means of manual inspection that may have errors. We make all slices available to third party inspection in the supplementary material.


Our algorithm is designed to slice single-thread deterministic programs, 
adaptations when that is not the case might not be trivial (e.g.,\cite{4658058}). Regarding our use of a static analysis, building an analysis with perfect recall is a known technological challenge \cite{10.1145/2644805}, thus as well as static slicers our algorithm may -in practice- miss statements that were relevant to the criteria node. In this sense, our technique inherits the limitations that static analysis might have in terms of scale. We understand that in some settings re-execution may be prohibitively expensive, for instance applications that perform heavy initialization (e.g., web servers, databases, mobile apps, etc) \vic{or exhibit non-deterministic behavior. In those respects, we believe that record and replay techniques \cite{ronsse1999recplay} may be a key technology and that re-execution should be performed during replay phase.}

 \vspace{-1em}
\section{Related Work}


Dynamic Slicing is a concept presented by Korel and Laski~\cite{korel1988dynamic} and Agrawal et.al.~\cite{agrawal1990dynamic} where they presented the idea of closure data and control dependencies for a given input. Following algorithms, sometimes redefine what they consider as a dynamic slice \cite{agrawal1990dynamic,zhang2003precise,korel1998dynamic,DBLP:conf/esec/GyimothyBF99} usually in terms of precision, slice executability \cite{xu2005brief,binkley1996program,binkley2006theoretical} or syntactically correspondence \cite{harman2003amorphous}. Approaches and tools that followed seminal Korel-Laski's concept explore different trade offs and optimizations to scale (e.g., \cite{mund2006efficient,wang2008dynamic,zhang2005cost,zhang2006dynamic,javaslicer,4026852,1541156}, etc.) but, to the best of our knowledge, all of them work based a single traced-execution. \vic{Moreover, to the best of our knowledge, all available tools that produce precise dynamic slices capture  memory reference information to extract relevant dependencies. Javaslicer~\cite{javaslicer,wang2008dynamic,wang2004using} and DrDebug~\cite{wang2014drdebug}  do an on-demand traversal of that information as proposed in \cite{zhang2003precise,wang2008dynamic}. Slicer4J~\cite{ahmed2021slicer4j} do full preprocessing \cite{zhang2003precise} by building a dependency graph \cite{agrawal1990dynamic} out of memory references.}

Observational slicing~\cite{binkley2014orbs,binkley2019comparison,lee2021observation}, unlike classical approaches, is not meant to keep in the slice statements that are relevant from the point of view of control or data dependence but, instead, to get an executable slice that produces the same results for the slicing criteria.  They do that by following a delete-and-observe paradigm, where a candidate slice is built by deleting lines of a program and then compiled and executed to check whether the observed behaviour (e.g., the output) is the same than the original program. Observational slicing is a promising concept designed mainly to deal with multiple languages since, by problem definition, it does not neither require dependency definitions nor ad-hoc instrumentation technology like it is the case for our and the tracing approaches. However, delete-and-observe requires a clever treatment of the potentially huge search space and, in some settings, every speculative deletion requires a compilation and then a successful a test execution \cite{lee2021observation}. Moreover deletion-and-observe techniques must cope with problems such as potential non-termination of slice candidates. Our approach instead do not impose a mandatory recompilation task, instead our observation can be done dynamically using the same program.

\vic{There is a few reported hybrid slicing approaches that blend static information with dynamic observation. ``Approach 2'' of~\cite{agrawal1990dynamic} is based on dynamically marking a static dependency graph. Interestingly, it has been reported as a rather imprecise approach~\cite{agrawal1990dynamic,zhang2005cost} and our upfront-all corroboration approach can be regarded as a more precise way to do such marking due to the contextualized nature of our corroboration infrastructure. Yet, our experiments have shown how this corroboration approach is less efficient than our on-demand approach because its performance is tightly coupled with execution size.}
Other reported hybrid approaches \cite{gupta1997hybrid,rilling2001hybrid} propose a different blend of static and dynamic information, for instance \cite{gupta1997hybrid} uses available dynamic information in debugging (manually introduced breakpoints) to prune the statically reported paths and build an static slice from them. Instead, our approach uses information of def-use coverage which is more precise than breakpoints information, which we can afford due to multiple re-executions.
Finally, we are aware of few dynamic analysis that executed more than once the program under analysis. Notably, secure multi-execution at different security-levels is used to enforce non-interference \cite{5504711}.
\vic {In \cite{DBLP:conf/icse/MadsenTASM16} a feedback-directed instrumentation technique
for computing crash paths that allows the instrumentation
overhead to be distributed over a crowd of users and to reduce it for users who do not encounter the crash. As authors noticed, this is a different problem than slicing and thus challenges and technical approaches are different in many aspects }

Record and replay technology (RR) focuses on providing low overhead recording to allow replaying program behaviour (e.g., \cite{10.5555/3154690.3154727,10.1145/1772954.1772958}). This is a related problem but not equivalent to identifying data flow dependencies in a program execution, our focus. Actually, tools supporting debugging workflows have made important progress by using record and replay technology but still have problems scaling with respect to execution size when (pre)computing and using information necessary for data flow tracking. For instance,  Pernosco \cite{pernosco} allows navigating data dependencies, for which it “pre-computes all program states” which can take minutes and that may fail when a submitted recording is ‘too large’. As mentioned, DrDebug provides dynamic slicing functionality on top of record replay technology. Particularly, memory reference information is captured during replay. As noted by \cite{wang2014drdebug} “both PinPlay (the RR) and dynamic slicing can incur in a large runtime overhead” that is why developer needs to manually specify ‘buggy region’ (which jeopardizes soundness of slices).

We believe that the ideas we present could be adopted by modern RR and debugging workflows to improve scalability of features aimed at supporting data flow dependencies analysis when traditional memory reference tracing do not scale.

\section{Conclusion and Future Work}

In this paper, we propose a novel approach that aims to offer an alternative to the classic execute-once paradigm built around Korel-Laski’s algorithm. The core idea is to re-execute the program to be sliced multiple times while keeping the need for tracking minimal and specific to the analysis needs at each execution. We present a concrete algorithm that follows the on-demand re-execution paradigm  that uses a novel concept of frontier dependency to incrementally build a dynamic slice. We show results of an evaluation on the SV-COMP benchmark and Antrl4 unit tests that provides evidence that on-demand re-execution can provide performance gains and a practical shift of complexity to slice size rather than execution size, thriving when slice size is small and execution size is large.

\bibliographystyle{IEEEtran}

\begin{thebibliography}{10}
\providecommand{\url}[1]{#1} \csname url@samestyle\endcsname
\providecommand{\newblock}{\relax}
\providecommand{\bibinfo}[2]{#2}
\providecommand{\BIBentrySTDinterwordspacing}{\spaceskip=0pt\relax}
\providecommand{\BIBentryALTinterwordstretchfactor}{4}
\providecommand{\BIBentryALTinterwordspacing}{\spaceskip=\fontdimen2\font
plus \BIBentryALTinterwordstretchfactor\fontdimen3\font minus
  \fontdimen4\font\relax}
\providecommand{\BIBforeignlanguage}[2]{{%
\expandafter\ifx\csname l@#1\endcsname\relax
\typeout{** WARNING: IEEEtran.bst: No hyphenation pattern has been}%
\typeout{** loaded for the language `#1'. Using the pattern for}%
\typeout{** the default language instead.}%
\else \language=\csname l@#1\endcsname \fi #2}}
\providecommand{\BIBdecl}{\relax} \BIBdecl

\bibitem{agrawal1990dynamic}
H.~Agrawal and J.~R. Horgan, ``Dynamic program slicing,''
\emph{ACM SIGPlan
  Notices}, vol.~25, no.~6, pp. 246--256, 1990.

\bibitem{korel1988dynamic}
B.~Korel and J.~Laski, ``Dynamic program slicing,''
\emph{Information
  processing letters}, vol.~29, no.~3, pp. 155--163, 1988.

\bibitem{li2020more}
X.~Li and A.~Orso, ``More accurate dynamic slicing for better
supporting
  software debugging,'' in \emph{2020 IEEE 13th International Conference on
  Software Testing, Validation and Verification (ICST)}.\hskip 1em plus 0.5em
  minus 0.4em\relax IEEE, 2020, pp. 28--38.

\bibitem{soremekun2021locating}
E.~Soremekun, L.~Kirschner, M.~B{\"o}hme, and A.~Zeller,
``Locating faults with
  program slicing: an empirical analysis,'' \emph{Empirical Software
  Engineering}, vol.~26, no.~3, pp. 1--45, 2021.

\bibitem{guo2018empirical}
A.~Guo, X.~Mao, D.~Yang, and S.~Wang, ``An empirical study on the
effect of
  dynamic slicing on automated program repair efficiency,'' in \emph{2018 IEEE
  International Conference on Software Maintenance and Evolution
  (ICSME)}.\hskip 1em plus 0.5em minus 0.4em\relax IEEE, 2018, pp. 554--558.

\bibitem{zhang2003precise}
X.~Zhang, R.~Gupta, and Y.~Zhang, ``Precise dynamic slicing
algorithms,'' in
  \emph{25th International Conference on Software Engineering, 2003.
  Proceedings.}\hskip 1em plus 0.5em minus 0.4em\relax IEEE, 2003, pp.
  319--329.

\bibitem{wang2008dynamic}
T.~Wang and A.~Roychoudhury, ``Dynamic slicing on java bytecode
traces,''
  \emph{ACM Transactions on Programming Languages and Systems (TOPLAS)},
  vol.~30, no.~2, pp. 1--49, 2008.

\bibitem{zhang2005cost}
X.~Zhang, R.~Gupta, and Y.~Zhang, ``Cost and precision tradeoffs
of dynamic
  data slicing algorithms,'' \emph{ACM Transactions on Programming Languages
  and Systems (TOPLAS)}, pp. 631--661, 2005.

\bibitem{zhang2006dynamic}
X.~Zhang, S.~Tallam, and R.~Gupta, ``Dynamic slicing long running
programs
  through execution fast forwarding,'' in \emph{Proceedings of the 14th ACM
  SIGSOFT international symposium on Foundations of software engineering},
  2006, pp. 81--91.

\bibitem{wang2014drdebug}
Y.~Wang, H.~Patil, C.~Pereira, G.~Lueck, R.~Gupta, and I.~Neamtiu,
``Drdebug:
  Deterministic replay based cyclic debugging with dynamic slicing,'' in
  \emph{Proceedings of annual IEEE/ACM international symposium on code
  generation and optimization}, 2014, pp. 98--108.

\bibitem{binkley2004survey}
D.~W. Binkley and M.~Harman, ``A survey of empirical results on
program
  slicing.'' \emph{Adv. Comput.}, vol.~62, no. 105178, pp. 105--178, 2004.

\bibitem{su2017survey}
T.~Su, K.~Wu, W.~Miao, G.~Pu, J.~He, Y.~Chen, and Z.~Su, ``A
survey on
  data-flow testing,'' \emph{ACM Computing Surveys (CSUR)}, vol.~50, no.~1, pp.
  1--35, 2017.

\bibitem{santelices2007efficiently}
R.~Santelices and M.~J. Harrold, ``Efficiently monitoring
data-flow test
  coverage,'' in \emph{Proceedings of the 22th IEEE ACM international
  conference on Automated software engineering}, 2007, pp. 343--352.

\bibitem{javaslicer}
C.~Hammacher, ``Design and implementation of an efficient dynamic
slicer for
  java,'' \emph{Saarland University, Nov}, 2008.

\bibitem{ahmed2021slicer4j}
K.~Ahmed, M.~Lis, and J.~Rubin, ``Slicer4j: a dynamic slicer for
java,'' in
  \emph{{ESEC/FSE} '21: 29th {ACM} Joint European Software Engineering
  Conference and Symposium on the Foundations of Software Engineering}.\hskip
  1em plus 0.5em minus 0.4em\relax {ACM}, 2021, pp. 1570--1574.

\bibitem{codeql}
``{CodeQL Tool},'' https://securitylab.github.com/tools/codeql/.

\bibitem{de2007ql}
O.~De~Moor, M.~Verbaere, E.~Hajiyev, P.~Avgustinov, T.~Ekman,
N.~Ongkingco,
  D.~Sereni, and J.~Tibble, ``Keynote address:. ql for source code analysis,''
  in \emph{7th IEEE International Working Conference on Source Code Analysis
  and Manipulation (SCAM 2007)}, pp. 3--16.

\bibitem{SV-COMP}
``{SV-COMP},''
  https://github.com/sosy-lab/sv-benchmarks/tree/master/java/algorithms.

\bibitem{antlr4}
``{Antlr4},'' https://github.com/antlr/antlr4.

\bibitem{wang2004using}
T.~Wang and A.~Roychoudhury, ``Using compressed bytecode traces
for slicing
  java programs,'' in \emph{Proceedings. 26th International Conference on
  Software Engineering}.\hskip 1em plus 0.5em minus 0.4em\relax IEEE, 2004, pp.
  512--521.

\bibitem{ferrante1987program}
J.~Ferrante, K.~J. Ottenstein, and J.~D. Warren, ``The program
dependence graph
  and its use in optimization,'' \emph{ACM Transactions on Programming
  Languages and Systems (TOPLAS)}, vol.~9, no.~3, pp. 319--349, 1987.

\bibitem{kang2011dta}
M.~G. Kang, S.~McCamant, P.~Poosankam, and D.~Song, ``Dta++:
dynamic taint
  analysis with targeted control-flow propagation.'' in \emph{NDSS}, 2011.

\bibitem{arzt2016stubdroid}
S.~Arzt and E.~Bodden, ``Stubdroid: automatic inference of precise
data-flow
  summaries for the android framework,'' in \emph{2016 IEEE/ACM 38th
  International Conference on Software Engineering (ICSE)}.\hskip 1em plus
  0.5em minus 0.4em\relax IEEE, 2016, pp. 725--735.

\bibitem{sridharan2011f4f}
M.~Sridharan, S.~Artzi, M.~Pistoia, S.~Guarnieri, O.~Tripp, and
R.~Berg, ``F4f:
  taint analysis of framework-based web applications,'' in \emph{Proceedings of
  the 2011 ACM international conference on Object oriented programming systems
  languages and applications}, 2011, pp. 1053--1068.

\bibitem{asm}
``{ASM},'' https://asm.ow2.io/.

\bibitem{jvmti}
``{Java Virtual Machine Tool Interface (JVM TI)},''
  https://docs.oracle.com/javase/8/docs/technotes/guides/jvmti.

\bibitem{yoo2012regression}
S.~Yoo and M.~Harman, ``Regression testing minimization, selection
and
  prioritization: a survey,'' \emph{Software testing, verification and
  reliability}, vol.~22, no.~2, pp. 67--120, 2012.

\bibitem{pythonresource}
``{Python3 resource library},''
  \url{https://docs.python.org/3/library/resource.html}.

\bibitem{ranganath2005new}
V.~P. Ranganath, T.~Amtoft, A.~Banerjee, M.~B. Dwyer, and
J.~Hatcliff, ``A new
  foundation for control-dependence and slicing for modern program
  structures,'' in \emph{European Symposium on Programming}.\hskip 1em plus
  0.5em minus 0.4em\relax Springer, 2005, pp. 77--93.

\bibitem{4658058}
S.~Tallam, C.~Tian, and R.~Gupta, ``Dynamic slicing of
multithreaded programs
  for race detection,'' in \emph{2008 IEEE International Conference on Software
  Maintenance}, 2008, pp. 97--106.

\bibitem{10.1145/2644805}
B.~Livshits, M.~Sridharan, Y.~Smaragdakis, O.~Lhot\'{a}k, J.~N.
Amaral,
  B.-Y.~E. Chang, S.~Z. Guyer, U.~P. Khedker, A.~M\o{}ller, and D.~Vardoulakis,
  ``In defense of soundiness: A manifesto,'' \emph{Commun. ACM}, vol.~58,
  no.~2, p. 44–46, 2015.

\bibitem{ronsse1999recplay}
M.~Ronsse and K.~De~Bosschere, ``Recplay: A fully integrated
practical
  record/replay system,'' \emph{ACM Transactions on Computer Systems (TOCS)},
  vol.~17, no.~2, pp. 133--152, 1999.

\bibitem{korel1998dynamic}
B.~Korel and J.~Rilling, ``Dynamic program slicing methods,''
\emph{Information
  and Software Technology}, vol.~40, pp. 647--659, 1998.

\bibitem{DBLP:conf/esec/GyimothyBF99}
T.~Gyim{\'{o}}thy, {\'{A}}.~Besz{\'{e}}des, and I.~Forg{\'{a}}cs,
``An
  efficient relevant slicing method for debugging,'' in \emph{Software
  Engineering - ESEC/FSE'99, 7th European Software Engineering Conference},
  vol. 1687.\hskip 1em plus 0.5em minus 0.4em\relax Springer, 1999, pp.
  303--321.

\bibitem{xu2005brief}
B.~Xu, J.~Qian, X.~Zhang, Z.~Wu, and L.~Chen, ``A brief survey of
program
  slicing,'' \emph{ACM SIGSOFT Software Engineering Notes}, vol.~30, no.~2, pp.
  1--36, 2005.

\bibitem{binkley1996program}
D.~W. Binkley and K.~B. Gallagher, ``Program slicing,''
\emph{Advances in
  computers}, vol.~43, pp. 1--50, 1996.

\bibitem{binkley2006theoretical}
D.~Binkley, S.~Danicic, T.~Gyim{\'o}thy, M.~Harman, {\'A}.~Kiss,
and B.~Korel,
  ``Theoretical foundations of dynamic program slicing,'' \emph{Theoretical
  Computer Science}, vol. 360, no. 1-3, pp. 23--41, 2006.

\bibitem{harman2003amorphous}
M.~Harman, D.~Binkley, and S.~Danicic, ``Amorphous program
slicing,''
  \emph{Journal of Systems and Software}, vol.~68, no.~1, pp. 45--64, 2003.

\bibitem{mund2006efficient}
G.~B. Mund and R.~Mall, ``An efficient interprocedural dynamic
slicing
  method,'' \emph{Journal of Systems and Software}, vol.~79, no.~6, pp.
  791--806, 2006.

\bibitem{4026852}
A.~Beszedes, T.~Gergely, and T.~Gyimothy, ``Graph-less dynamic
dependence-based
  dynamic slicing algorithms,'' in \emph{2006 6th IEEE International Workshop
  on Source Code Analysis and Manipulation}, 2006, pp. 21--30.

\bibitem{1541156}
A.~Szegedi and T.~Gyimothy, ``Dynamic slicing of java bytecode
programs,'' in
  \emph{5th IEEE International Workshop on Source Code Analysis and
  Manipulation (SCAM'05)}, 2005, pp. 35--44.

\bibitem{binkley2014orbs}
D.~Binkley, N.~Gold, M.~Harman, S.~Islam, J.~Krinke, and S.~Yoo,
``Orbs:
  Language-independent program slicing,'' in \emph{Proceedings of the 22nd ACM
  SIGSOFT International Symposium on Foundations of Software Engineering},
  2014, pp. 109--120.

\bibitem{binkley2019comparison}
D.~Binkley, N.~Gold, S.~Islam, J.~Krinke, and S.~Yoo, ``A
comparison of
  tree-and line-oriented observational slicing,'' \emph{Empirical Software
  Engineering}, vol.~24, no.~5, pp. 3077--3113, 2019.

\bibitem{lee2021observation}
S.~Lee, D.~Binkley, R.~Feldt, N.~Gold, and S.~Yoo,
``Observation-based
  approximate dependency modeling and its use for program slicing,''
  \emph{Journal of Systems and Software}, vol. 179, p. 110988, 2021.

\bibitem{gupta1997hybrid}
R.~Gupta, M.~L. Soffa, and J.~Howard, ``Hybrid slicing:
Integrating dynamic
  information with static analysis,'' \emph{ACM Transactions on Software
  Engineering and Methodology (TOSEM)}, vol.~6, no.~4, pp. 370--397, 1997.

\bibitem{rilling2001hybrid}
J.~Rilling and B.~Karanth, ``A hybrid program slicing framework,''
in
  \emph{Proceedings First IEEE International Workshop on Source Code Analysis
  and Manipulation}.\hskip 1em plus 0.5em minus 0.4em\relax IEEE, 2001, pp.
  12--23.

\bibitem{5504711}
D.~Devriese and F.~Piessens, ``Noninterference through secure
  multi-execution,'' in \emph{2010 IEEE Symposium on Security and Privacy},
  2010, pp. 109--124.

\bibitem{DBLP:conf/icse/MadsenTASM16}
M.~Madsen, F.~Tip, E.~Andreasen, K.~Sen, and A.~M{\o}ller,
``Feedback-directed
  instrumentation for deployed javascript applications,'' in \emph{Proceedings
  of the 38th International Conference on Software Engineering, {ICSE}}.\hskip
  1em plus 0.5em minus 0.4em\relax {ACM}, 2016, pp. 899--910.

\bibitem{10.5555/3154690.3154727}
R.~O'Callahan, C.~Jones, N.~Froyd, K.~Huey, A.~Noll, and
N.~Partush,
  ``Engineering record and replay for deployability,'' in \emph{Proceedings of
  the 2017 USENIX Conference on Usenix Annual Technical Conference}.\hskip 1em
  plus 0.5em minus 0.4em\relax USENIX, 2017, p. 377–389.

\bibitem{10.1145/1772954.1772958}
H.~Patil, C.~Pereira, M.~Stallcup, G.~Lueck, and J.~Cownie,
``Pinplay: A
  framework for deterministic replay and reproducible analysis of parallel
  programs,'' in \emph{Proceedings of the 8th Annual IEEE/ACM International
  Symposium on Code Generation and Optimization}, ser. CGO '10.\hskip 1em plus
  0.5em minus 0.4em\relax NY, USA: ACM, 2010, p. 2–11.

\bibitem{pernosco}
``{Pernosco},'' https://pernos.co/.

\end{thebibliography}



%

\clearpage
\clearpage

\ifCLASSOPTIONcaptionsoff
  \newpage
\fi

\end{document}